\documentclass[times]{aastex631}
\usepackage{amssymb}
\usepackage{amsmath}
\usepackage{gensymb}
\usepackage{graphicx}
\usepackage{xcolor}
\usepackage{natbib}

\newcommand{\beq}{\begin{equation}}
\newcommand{\eeq}{\end{equation}}

\newcommand{\Ms}{\textrm{M}_*}
\newcommand{\Msun}{\textrm{M}_\odot}
\newcommand{\kmps}{km~s$^{-1}$}
\newcommand{\MHI}{\rm{M_{H{\textsc i}}}}
\newcommand{\Mat}{\rm{M_{Atom}}}

\newcommand{\htwo}{${\rm H_2}$}

\newcommand{\hi}{H{\sc i}}

\newcommand{\hii}{H{\sc i} 21\,cm}

\newcommand{\Mmol}{\rm{M_{Mol}}}

\newcommand{\MBary}{\rm{M_{Baryon}}}

\defcitealias{Chowdhury22a}{C22a}
\defcitealias{Chowdhury22b}{C22b}

\graphicspath{{./}{figures/}}

\shorttitle{Cold Gas Content at $z\approx1$}
\shortauthors{Chowdhury, Kanekar and Chengalur}

\begin{document}

	\title{Atomic Gas Dominates the Baryonic Mass of Star-forming Galaxies at $z \approx 1.3$}

	\correspondingauthor{Nissim Kanekar}
	\email{nkanekar@ncra.tifr.res.in}
	
	\author{Aditya Chowdhury}
	\affil{National Centre for Radio Astrophysics, Tata Institute of Fundamental Research, Pune, India.}
	
	\author{Nissim Kanekar}
	\affil{National Centre for Radio Astrophysics, Tata Institute of Fundamental Research, Pune, India.}
	
	\author{Jayaram N. Chengalur}
	\affil{National Centre for Radio Astrophysics, Tata Institute of Fundamental Research, Pune, India.}

	
	
	\begin{abstract}
We present a comparison between the average atomic gas mass, $\rm \langle M_{\rm Atom} \rangle$ {(including hydrogen and helium)}, the average molecular gas mass, $\rm \langle M_{\rm Mol} \rangle$, and the average stellar mass, $\rm \langle M_\star \rangle$, of a sample of star-forming galaxies at $z \approx 0.75-1.45$, to probe the baryonic composition of galaxies in and during the epoch of peak star-formation activity in the universe.
The $\rm \langle M_{\rm Atom} \rangle$ values of star-forming galaxies in two stellar-mass matched samples at $z=0.74-1.25$ and $z=1.25-1.45$, were derived by stacking their H{\sc i} 21\,cm signals in the GMRT-CAT$z1$ survey. We find that the baryonic composition of star-forming galaxies at $z \gtrsim 1$ is dramatically different from that at $z \approx 0$. For star-forming galaxies with  $\langle \textrm{M}_*\rangle \approx10^{10}~\textrm{M}_\odot$, the contribution of stars to the total baryonic mass, ${\rm M_{Baryon}}$,  is $\approx 61\%$ at $z \approx 0$, but only $\approx 16$\% at $z \approx 1.3$, while molecular gas constitutes $\approx6\%$ of the baryonic mass at $z\approx0$, and  $\approx14\%$ at $z\approx1.3$. {Remarkably, we find that atomic gas makes up $\approx70\%$ of ${\rm M_{Baryon}}$ in star-forming galaxies at $z\approx1.3$.  We find that the ratio $\rm \langle M_{\rm Atom} \rangle/\langle M_\star \rangle$ is higher both at $z\approx1.0$ and at $z\approx1.3$ than in the local Universe, with $\rm \langle M_{\rm Atom} \rangle/\langle M_\star \rangle \approx 1.4$ at $z \approx 1.0$, and $\approx4.4$ at $z\approx1.3$, compared to its value of $\approx 0.5$ today. Further, we find that the ratio $\rm \langle M_{\rm Atom} \rangle/\langle M_{\rm Mol} \rangle$ in star-forming galaxies with $\langle \textrm{M}_*\rangle \approx10^{10}~\textrm{M}_\odot$ is $\approx 2.3$ at $z \approx 1.0$ and $\approx 5.0$ at $z\approx1.3$.} Overall, we find that atomic gas is the dominant component of the baryonic mass of star-forming galaxies at $z\approx 1.3$, during the epoch of peak star-formation activity in the universe.

	\end{abstract}
	
	
\section{Introduction}

Neutral atomic hydrogen (\hi) and molecular hydrogen (\htwo) are the key components of the cold interstellar medium (ISM) in galaxies, and are the fuel for star formation. The distribution of baryons between atomic gas, molecular gas, and stars is an important indicator of the evolutionary stage of a galaxy: at early times, most of the baryonic mass is in the atomic phase, while, for highly evolved systems (e.g. red and dead ellipticals), almost all the baryons are in the stars. The relative contributions of atomic gas, molecular gas, and stars to the total baryonic mass in galaxies, and the evolution of these contributions over cosmological time are thus critical inputs to studies of galaxy evolution. In the local Universe, most of the baryonic content in massive star-forming galaxies at $z\approx0$ is in stars, while \hi\ makes up $\approx 70-90$\% of the cold-gas content of most such galaxies \citep[e.g.][]{Saintonge17,Catinella18}.

At high redshifts, CO observations of star-forming galaxies at $z \approx 1 - 3$ have found evidence for large reservoirs of molecular gas, comparable in mass to their stellar masses  \citep[e.g.][]{Daddi10,Tacconi13,Tacconi18}. This is very different from the situation in galaxies at $z \approx 0$, where the ratio of the molecular gas mass to the stellar mass is only $\approx 0.02-0.10$ \citep{Saintonge17}. Indeed, the molecular gas mass of main-sequence galaxies has been observed to increase by approximately an order of magnitude from $z\approx0$ to $z\approx2$ \citep[e.g.][]{Genzel15,Tacconi20}. The large inferred molecular gas content of high-$z$ galaxies has been used to argue that the cold-gas content of galaxies at $z\approx1-3$ is predominantly of molecular form \citep[e.g.][]{Tacconi18}. 

Measurements of the atomic and molecular content of high-$z$ galaxies provide important constraints on numerical and semi-analytical models of galaxy evolution  \citep[e.g.][]{Obreschkow09,Lagos11,Popping14,Dave19,Dave20}. Unfortunately, the weakness of the \hii\ line, the only tracer of the \hi\ content of galaxies, has made it very challenging to directly measure the \hi\ mass of galaxies at cosmological distances. Indeed, even at intermediate redshifts, there are only a handful of galaxies at $z\approx0.2-0.4$ for which estimates of $\Mat/\Mmol$ are available \citep{Cybulski16,Fernandez16,Cortese17}. As a result, it has hitherto not been possible to  measure the redshift evolution of $\Mat/\Mmol$ and directly test the hypothesis that most of the cold gas in star-forming galaxies at $z \approx 1-3$ is in the molecular phase.

Recently, the \hii\ stacking approach has been  used to measure the {\it average} \hi\ mass of star-forming galaxies out to $z\approx1.3$ \citep{Bera19,Chowdhury20,Chowdhury21}. \citet{Chowdhury20} used the upgraded Giant Metrewave Radio Telescope (GMRT) to measure, for the first time, the average \hi\ mass of galaxies at $z\approx1$, by stacking the \hii\ emission signals of 7,653 blue star-forming galaxies at $z=0.74-1.45$ in the DEEP2 survey fields \citep{Newman13}. More recently,  \citet[][hereafter, \citetalias{Chowdhury22a}]{Chowdhury22a}  used the GMRT Cold-\hi\ AT $z\approx1$ (CAT$z1$) survey (\citealp{Chowdhury22b}; hereafter, \citetalias{Chowdhury22b}), a 510-hr upgraded GMRT \hii\ emission survey of galaxies at $z=0.74-1.45$, also in the DEEP2 survey fields, to measure the average \hi\ mass of star-forming galaxies in two stellar-mass-matched subsamples at $z=0.74-1.25$ and $z=1.25-1.45$. \citetalias{Chowdhury22a} found that the average \hi\ mass of main-sequence galaxies declines steeply from $z\approx1.3$ to $z\approx1.0$, by a factor of $3.2\pm0.8$.  
    
In this \emph{Letter}, we combine the GMRT-CAT$z1$ measurements of the average \hi\ mass of star-forming galaxies at $z\gtrsim1$ with estimates of the average \htwo\ mass and the average stellar mass of the same galaxies, to estimate, for the first time, the contribution of atomic gas, molecular gas, and stars to the baryonic mass of galaxies at $z \gtrsim 1$, nearly nine billion years ago.

{Throughout this \emph{Letter}, we use a flat Lambda-cold dark matter cosmology, with $\Omega_m=0.3$, $\Omega_\Lambda = 0.7$, and $H_0 = 70$~\kmps~Mpc$^{-1}$. All estimates of stellar masses and SFRs assume a Chabrier initial mass function (IMF); stellar masses and SFRs from the literature that assume a Salpeter IMF were converted to a Chabrier IMF by subtracting 0.2~dex \citep[e.g.][]{Madau14}. }

\section{The cold gas content of galaxies at $z \approx 1$ and in the local Universe}
\label{sec:analysis}

\subsection{Atomic Gas in Star-forming Galaxies at $z=0.74-1.45$}
\begin{figure}
    \centering
    \includegraphics[width=\linewidth]{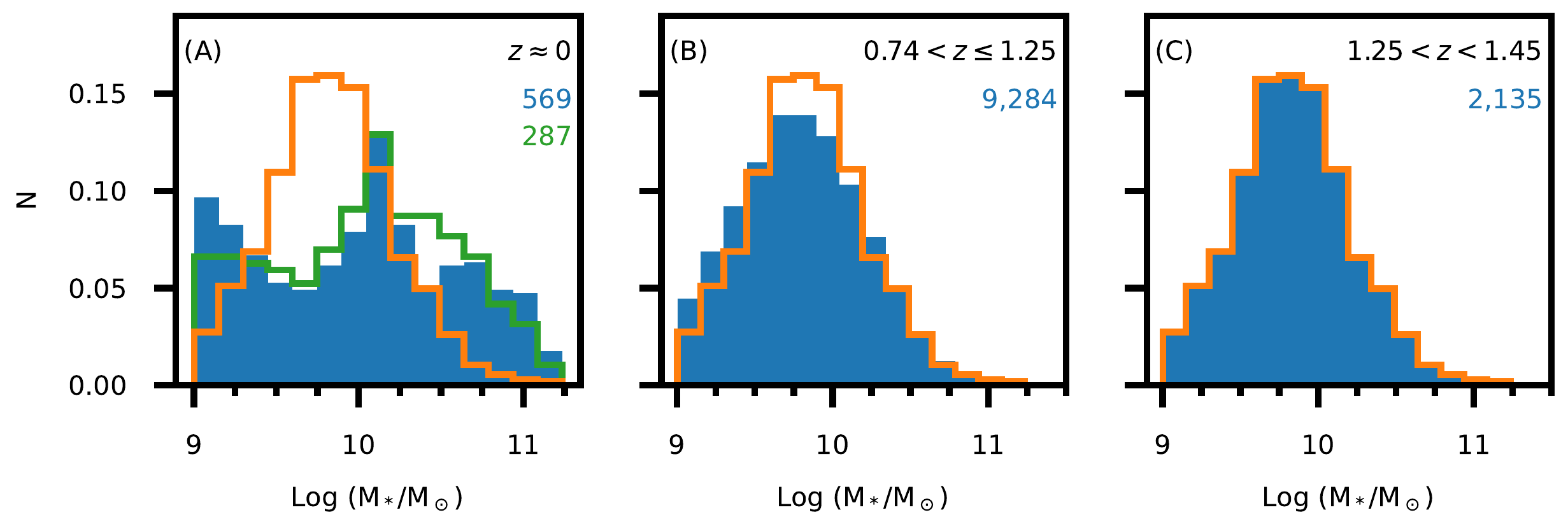}
    \caption{The stellar mass distribution of the galaxies in the three redshift intervals. Panel~(A) shows the stellar mass distribution of the blue galaxies of the xGASS sample \citep[blue histogram;][]{Catinella18} and those of the xCOLD~GASS sample \citep[green histogram;][]{Saintonge17}. Panels~(B) and (C) show the stellar-mass distributions of the GMRT-CAT$z1$ subsamples at $z=0.74-1.25$ and $z=1.25-1.45$ \citepalias[blue histograms;][]{Chowdhury22a}. All average quantities reported in this work, for the three subsamples at $z\approx0$, $z\approx1.0$, and $z\approx1.3$, were computed with weights such that the  stellar-mass distribution of each subsample is identical to that of the subsample at $z=1.25-1.45$ (the orange histogram in each panel). The number of galaxies in each subsample is listed in each panel. }
    \label{fig:stellar-mass}
\end{figure}
\begin{figure}
    \centering
    \includegraphics[width=\linewidth]{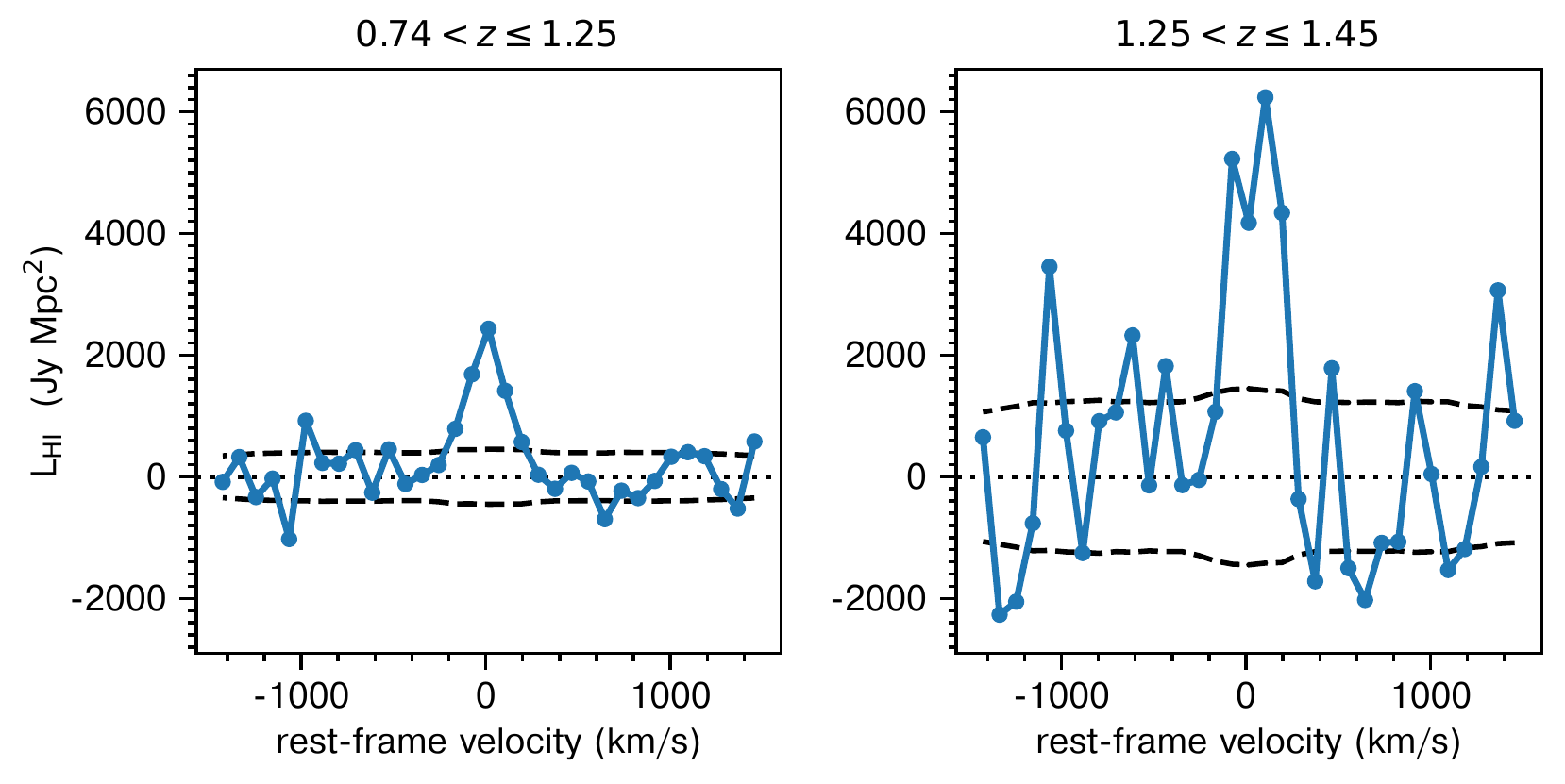}
    \caption{The stacked \hii\ spectra of blue star-forming galaxies at $z=0.74-1.25$ (left panel) and $z=1.25-1.45$ (right panel) from \citetalias{Chowdhury22b}. The stacked \hii\ spectrum of each panel was obtained by stacking the individual \hii\ spectra of the galaxies in the subsample, using weights to ensure that the stellar-mass distributions of the two subsamples are identical. The dashed curve in each panel shows the $1\sigma$ RMS noise error on each 90~\kmps\ velocity channel of the stacked \hii\ spectrum. }
    \label{fig:HIstacks}
\end{figure}

The GMRT-CAT$z1$ survey provides measurements of the average \hi\ mass of star-forming galaxies in two stellar-mass matched subsamples at $z=0.74-1.25$ and $z=1.25-1.45$  \citepalias{Chowdhury22a}. The \hii\ stacking analysis used to obtain the average \hi\ mass estimates is described in detail in \citetalias{Chowdhury22a}. We provide here a summary of relevant information on the sample of galaxies, and the \hii\ stacking analysis and results, for the two redshift intervals. 

The main sample of the GMRT-CAT$z1$ survey contains 11,419 blue star-forming galaxies with $\Ms\geq10^{9}~\Msun$ at $z=0.74-1.45$ in seven GMRT pointings on the DEEP2 survey fields \citepalias{Chowdhury22b}.  The stellar masses of the 11,419 galaxies were inferred from their rest-frame U$-$B colours, rest-frame B$-$V colours and rest-frame absolute B-band magnitudes  \citep{Weiner09}; the relation was calibrated via comparisons with DEEP2 galaxies at similar redshifts in regions with K-band photometry \citep{Weiner09}. {The SFRs of the 11,419 galaxies of our sample were inferred using a calibration from \citet{Mostek12}, based on their rest-frame B-band magnitudes and the rest-frame (U-B) colours.\footnote{This SFR calibration was derived by \citet{Mostek12} for DEEP2 galaxies in the Extended Groth Strip for which SFRs were obtained by \citet{Salim09} via spectral-energy distribution (SED) fits to the ultraviolet, optical, and near-infrared photometry. \citet{Salim09} found the SFRs inferred from the SED fits to be consistent with the mid-infrared luminosities of the DEEP2 galaxies.}} Dividing the galaxies into multiple redshift and stellar-mass bins, the average stellar masses and the average SFRs of galaxies in each bin are found to be consistent with the star-forming main sequence \citep{Whitaker14} at these redshifts \citepalias{Chowdhury22a}.

The GMRT-CAT$z1$ survey provides \hii\ subcubes for the 11,419 galaxies at a spatial resolution of 90~kpc and a velocity resolution of 90~\kmps\ \citepalias{Chowdhury22b}. {The subcubes of each galaxy were converted from flux density} ($\rm{S_{H{\textsc i}}}$, { in units of Jy) to luminosity density } ($\rm{L_{H{\textsc i}}}$, { in units of  Jy~Mpc$^2$) using the relation} $\rm{L_{H{\textsc i}} =4\pi~S_{H{\textsc i}}~D_L^2/(1+z)}$, { where ${\rm d_L}$ is the luminosity distance of the galaxy, in Mpc.} The spatial resolution of 90~kpc was chosen to ensure that the average \hii\ emission from the full sample of 11,419 galaxies is spatially unresolved \citepalias{Chowdhury22b}. 

The measurement of the average \hi\ mass of galaxies in the two redshift bins was obtained by dividing the 11,419 galaxies into two redshift subsamples, with $z=0.74-1.25$ (9284 galaxies) and $z=1.25-1.45$ (2135 galaxies), and separately stacking the \hii\ subcubes of the galaxies in each subsample \citepalias{Chowdhury22a}. The effective stellar-mass distributions of the two redshift subsamples were made identical by applying weights to the galaxies of the lower-$z$ sample during the stacking procedure; { the stellar-mass distributions of the two redshift subsamples are shown in Fig.~\ref{fig:stellar-mass}.} The average stellar mass of the subsample at $z\approx1.3$ is $\approx10^{10}~\Msun$. The stacked \hii\ spectral cube of each subsample of galaxies was then obtained by using the above weights to take a weighted average of the \hii\ subcubes of the DEEP2 galaxies in each subsample. { The RMS noise on each of the stacked \hii\ spectral cubes was estimated using Monte Carlo simulations, taking into account the stellar-mass based weights of the galaxies in each subsample \citepalias{Chowdhury22a}.} We note that the final stacked GMRT \hii\ spectral cubes have a spatial resolution of 90~kpc. The compact GMRT beam ensures that the measurements of the average \hi\ mass of galaxies in the GMRT-CAT$z1$ survey are not significantly affected by \hii\ emission from companion galaxies around the target galaxies, i.e. by source confusion \citepalias{Chowdhury22b}.

The average \hi\ masses of the galaxies in the two redshift subsamples were derived from their stacked \hii\ subcubes using the following procedure: (i)~the central velocity channels of the stacked cube were averaged to obtain a stacked \hii\ emission image of the subsample, (ii)~the \hii\ spectrum at the location of the peak luminosity-density in the stacked \hii\ emission image was extracted, (iii)~contiguous central velocity channels of the stacked spectrum, with emission detected at $> 1.5\sigma$ statistical significance, were integrated to measure the average velocity-integrated \hii\ line luminosity ($\int \rm{L_{H{\textsc i}} \ dV}$, in units of ${\rm Jy~Mpc^2}$~\kmps) of the subsample, and (iv)~the average velocity-integrated line luminosity was converted to the average \hi\ mass of the subsample via the relation  $\MHI=[1.86 \times 10^4 \times \int \rm{L_{H{\textsc i}} \ dV}]~\Msun$.

The stacked \hii\ emission spectra of the 9284 galaxies at $z = 0.74-1.25$ and the 2135 galaxies at $z=1.25-1.45$ are shown in Fig.~\ref{fig:HIstacks}. For both redshift subsamples, the stacked \hii\ emission signal is clearly detected, at $>5.2~\sigma$ statistical significance. { We convert the measured average \hi\ mass of galaxies in each subsample to an estimate of the average atomic gas mass of the subsample using the relation} $\langle\Mat\rangle=1.36 \times \langle \MHI \rangle$, { where the factor of 1.36 accounts for the mass contribution of helium.}  The average atomic gas masses of the galaxies in the two subsamples are listed in Table~\ref{tab:mass}.

\subsection{Atomic Gas in Star-forming Galaxies at $z\approx0$}
We use the extended GALEX Arecibo SDSS Survey \citep[xGASS;][]{Catinella18} of nearby galaxies as a reference sample, to compare the \hi\ properties of galaxies at $z\gtrsim1$ to those of galaxies in the local Universe. xGASS is an \hii\ survey of a stellar-mass selected sample of local-Universe galaxies with $\Ms>10^{9} \ \Msun$ \citep{Catinella18}. Each galaxy in the xGASS sample was observed with the Arecibo Telescope until either a detection of the \hii\ emission was obtained or a $3\sigma$ upper limit of $\leq 0.1$ was achieved on the ratio of the \hi\ mass to the stellar mass. In order to carry out a fair comparison with our sample of blue star-forming galaxies at $z\approx1$, we used only the 569 blue xGASS galaxies, with $\rm NUV - r < 4$, as the reference sample. The stellar masses of all the xGASS galaxies are available from the Sloan Digital Sky Survey DR7 MPA-JHU catalog \citep{Kauffmann03,Brinchmann04}; { Figure~\ref{fig:stellar-mass}[A] shows the stellar mass distribution of the 569 blue xGASS galaxies.}

We measured the average \hi\ mass of the blue xGASS galaxies, using weights such that the stellar-mass distribution of the blue xGASS sample is identical to that of our DEEP2 galaxies at $z=1.25-1.45$.  {  The \hii\ line was not detected for 16 of the 569 blue galaxies; for these 16 galaxies, we assume that the \hi\ mass is equal to the $3\sigma$ upper limit on} $\MHI$. { The error on the average \hi\ mass was estimated using bootstrap resampling with replacement. We note that we used the same stellar-mass based weights during the bootstrap resampling procedure in order to compute the weighted-average \hi\ mass of each randomly-drawn subsample. Finally, we again converted the average \hi\ mass of the sample to the average atomic gas mass via the relation} $\langle\Mat\rangle=1.36 \times \langle\MHI\rangle$.  Table~\ref{tab:mass} lists the average atomic gas mass of the blue xGASS galaxies at $z\approx 0$, with $\langle \Ms\rangle\approx10^{10}~\Msun$. 

\subsection{Molecular Gas in Star-forming Galaxies at $z=0.74-1.45$}

The H$_2$ mass of galaxies is typically estimated from tracers of molecular gas, such as the CO rotational lines, the far-infrared dust continuum, or the 1-mm dust continuum \citep[e.g.][]{Tacconi20}, with the different methods based on different assumptions and calibration schemes.
 \citet{Tacconi20} used a compilation of molecular gas mass estimates from the literature (including the mass contribution from helium) to provide the following relation between the molecular gas depletion timescale ($\rm t_{dep;mol}=M_{Mol}$/SFR) of galaxies and their (i)~redshift, (ii)~stellar mass, and (iii)~offset from the star-forming main-sequence at the galaxy redshift:

\begin{equation}
    \log[{\rm t_{dep;mol}/Gyr}] = {\rm A} - {\rm B} \times \log[1 + z] + {\rm C} \times \log[{\rm sSFR}/{\rm sSFR}({\rm MS}, z, \Ms )] + {\rm D} \times (\log[\Ms/\Msun] - 10.7)\ \ ,
    \label{eqn:tacconi20}
\end{equation} 

\noindent where sSFR ($\rm \equiv SFR/\Ms$) is the specific star-formation rate, ${\rm sSFR}({\rm MS}, z, \Ms )$ is the sSFR of galaxies with stellar mass $\Ms$  lying on the star-forming main-sequence at redshift $z$, and the values of the co-coefficients are A$=0.21\pm0.10$, B$=-0.98\pm0.10$, C$=-0.49\pm0.03$, and D$=0.03\pm0.04$ \citep[where the quoted errors are $2\sigma$ uncertainties; ][]{Tacconi20}. The relation was obtained from a sample of $2052$ galaxies with redshifts $z \approx 0 - 5.3$, stellar masses $\Ms \approx 10^9 - 10^{12.2} \ \Msun$, and SFRs~$\approx 0.04 - 5600 \ \Msun$~yr$^{-1}$ \citep{Tacconi20}. 

We use Equation~\ref{eqn:tacconi20} to estimate the molecular  gas depletion timescale of each of the 11,419 GMRT-CAT$z1$ galaxies. Next, we combine the molecular depletion timescales of the individual galaxies with their SFRs to infer the molecular gas  mass of each galaxy. Finally, we take a weighted mean of the $\Mmol$ values in the two redshift subsamples, with the same weights (see Fig.~\ref{fig:stellar-mass}) that were used while stacking the \hii\ emission from each subsample.
The estimated average molecular gas masses of the galaxies in the two redshift subsamples are listed in Table~\ref{tab:mass}.

{ The errors on the coefficients A, B, C, and D were propagated via a Monte Carlo approach to estimate the formal error on the average molecular gas mass of each subsample, appropriately taking into account the weight associated with each galaxy in the subsample}; these errors are listed in Table~\ref{tab:mass}.  \citet{Tacconi20} note that systematic uncertainties in reduced quantities like the sSFR have little effect on the inferred ${\rm t_{dep;mol}}$. This is even more the case for our sample of DEEP2 galaxies, which lie on the main sequence \citepalias{Chowdhury22a}; we have further verified that even excluding the sSFR dependence from Equation~\ref{eqn:tacconi20} has no significant effect on the average molecular gas mass.
However, we note that the formal error on the average molecular gas mass for each subsample does not include uncertainties stemming from the assumptions (e.g. the CO-to-H$_2$ conversion factor, $\alpha_{\rm CO}$) made in the molecular gas mass estimates of the original sample of 2052 galaxies. \citet{Tacconi20} estimate that the uncertainty arising from the assumptions is $\approx 0.25$~dex; the effect of these uncertainties is discussed in Section~\ref{sec:discuss}.

Although Equation~\ref{eqn:tacconi20} was obtained from a sample of galaxies with $\Ms \approx 10^9 - 10^{12.2} \ \Msun$ and at $z \approx 0-5.3$, the vast majority of $\Mmol$ estimates in galaxies at $z \gtrsim 0.5$ are for objects with $\Ms \gtrsim 10^{10} \ \Msun$ \citep[e.g.][]{Tacconi20}. The stellar mass of the DEEP2 galaxies in the GMRT CAT$z$1 survey extends down to $10^9 \ \Msun$ at $z \approx 0.74 - 1.45$ \citepalias{Chowdhury22a}, implying that we are applying the relation of \citet{Tacconi20} in a regime where it is not well constrained.

However, an alternative way to estimate the molecular gas masses of the DEEP2 galaxies is from the molecular gas depletion timescale, which is $\approx 0.7$~Gyr in main-sequence galaxies at $z\approx1$, with only a weak dependence on the stellar mass \citep{Tacconi13,Genzel15}. Assuming a constant molecular gas depletion timescale of $0.7$~Gyr, we find that the inferred average $\Mmol$ values for the GMRT-CAT$z1$ galaxies are consistent with those obtained from Equation~\ref{eqn:tacconi20}. It is hence unlikely that our estimates of the average $\Mmol$ of star-forming galaxies at $z\approx1.0$ and $z\approx1.3$ are significantly affected by the above extrapolation to lower stellar masses.

\begin{table}[]
    \centering
    \begin{tabular}{|l|c|c|c|}
        \hline\hline
         & $z\approx0$ & $z\approx1$ & $z\approx1.3$  \\
         \hline
       Average Stellar Mass, $\langle\Ms\rangle$ ($10^9~\Msun$) & $10.3 \pm 2.4$ & $10.3 \pm 2.4$ & $10.3 \pm 2.4$  \\
       Average Atomic Gas Mass, $\langle \Mat \rangle$ ($10^9~\Msun$) & $5.59\pm0.24$ & $14.4\pm2.6$ & $45.7\pm8.7$ \\
      Average Molecular Gas Mass, $\langle \Mmol \rangle$ ($10^9~\Msun$) & $0.974\pm0.048$ & $6.22\pm0.78$ & $9.19\pm1.17$ \\
     Average Baryonic Mass, $\langle \textrm{M}_\textrm{Baryon} \rangle$ ($10^9~\Msun$) & $16.86\pm2.4$ & $30.9\pm3.6$ & $65.3\pm9.1$ \\
        \hline 
        \hline
    \end{tabular}
    
    \caption{The average masses of the key baryonic constituents of stellar-mass matched samples of blue star-forming galaxies at three different redshifts, $z \approx 0$, $z \approx 1$, and $z \approx 1.3$. The four rows list, for each redshift, (1)~the average stellar mass, $\langle\Ms\rangle$, (2)~the average atomic gas mass, $\langle\Mat\rangle$, including the mass contribution from helium, (3)~the average molecular gas mass, $\langle\Mmol\rangle$, again including the mass contribution from helium, and (4)~the total average baryonic mass,  $\langle\MBary\rangle\equiv\langle\Ms\rangle+\langle\Mat\rangle+\langle\Mmol\rangle$. 
    The average atomic gas masses of blue star-forming galaxies at $z\approx1.0$ and $z\approx1.3$ are from the GMRT-CAT$z1$ survey \citepalias{Chowdhury22a}, with the average molecular gas masses of the same galaxies estimated using Equation~\ref{eqn:tacconi20} \citep{Tacconi20}. The average atomic gas and molecular gas masses of blue star-forming galaxies at $z\approx0$ were obtained, respectively, from the xGASS and xCOLD~GASS surveys \citep{Saintonge17,Catinella18}.  The errors indicate the $1\sigma$ uncertainties in the estimates. See text for discussion.}
    \label{tab:mass}
    \end{table}
      
\begin{table}[]
\centering
\begin{tabular}{|l|c|c|c|}
\hline\hline
    & $z\approx0$ & $z\approx1$ & $z\approx1.3$  \\
    \hline
    & & & \\
    Average atomic-gas-to-stars mass ratio,    $\langle \Mat \rangle$/$\langle\Ms\rangle$ & $0.54^{+0.17}_{-0.10}$ & $1.40^{+0.51}_{-0.35}$ & $4.44^{+1.65}_{-1.14}$ \\[0.15cm] 
    Average molecular-gas-to-stars mass ratio,     $\langle \Mmol \rangle$/$\langle\Ms\rangle$  & $0.095^{+0.029}_{-0.018}$ & $0.60^{+0.20}_{-0.13}$ & $0.89^{+0.30}_{-0.20}$ \\[0.25cm] 
       \hline 
       & & &\\
    Average atomic-gas-to-baryons mass ratio,   $\langle \Mat \rangle$/$\langle \textrm{M}_\textrm{Baryon} \rangle$ $\times10^2$ & $33.1^{+5.6}_{-4.2}$ & $46.6^{+5.8}_{-6.0}$ & $70.1^{+4.6}_{-5.4}$ \\[0.15cm] 
     Average molecular-gas-to-baryons mass ratio,   $\langle \Mmol \rangle$/$\langle \textrm{M}_\textrm{Baryon} \rangle$ $\times10^2$ & $5.78^{+1.01}_{-0.77}$ & $20.1^{+3.3}_{-2.9}$ & $14.1^{+2.8}_{-2.3}$ \\[0.15cm] 
     Average stars-to-baryons mass ratio,    $\langle\Ms\rangle$/$\langle \textrm{M}_\textrm{Baryon} \rangle$ $\times10^2$ & $61.1^{+6.5}_{-4.9}$ & $33.3^{+6.1}_{-5.8}$ & $15.8^{+3.6}_{-4.0}$ \\[0.25cm] 
         \hline 
         & & &\\
     Average atomic-to-molecular gas mass ratio,    $\langle \Mat \rangle$/$\langle \Mmol \rangle$  & $5.74^{+0.39}_{-0.37}$ & $2.32^{+0.54}_{-0.47}$ & $5.0^{+1.2}_{-1.0}$ \\[0.25cm] 
        \hline
        \hline
    \end{tabular}
    \caption{The ratios of the average masses of the key baryonic constituents of blue star-forming galaxies at three different redshifts, $z \approx 0$, $z \approx 1$, and $z \approx 1.3$. The first two rows list the ratios of the average atomic gas mass and the average molecular gas mass to the average stellar mass. The next three rows list the ratios of the average atomic gas mass, the average molecular gas mass, and the average stellar mass to the average total baryonic mass. Finally, the last row lists the ratio of the average atomic gas mass to the average molecular gas mass.  The errors indicate the $1\sigma$ uncertainties in the estimates. See text for discussion. }
    \label{tab:ratios}
\end{table}

\subsection{Molecular Gas in Star-forming Galaxies at $z\approx0$}

  We use the extended CO Legacy Database for GASS \citep[xCOLD~GASS][]{Saintonge17} survey to compute the average molecular gas mass of a reference sample of blue star-forming galaxies at $z\approx0$. The xCOLD~GASS survey used the IRAM~30m telescope to carry out CO(1--0) observations of a sample of 532 galaxies at $z \approx 0.01-0.05$ and with stellar masses $\Ms > 10^9 \ \Msun$ \citep{Saintonge17}. Approximately $90$\% of the xCOLD~GASS galaxies are covered in the \hii\ line with the xGASS survey \citep{Saintonge17,Catinella18}, while stellar masses for all xCOLD~GASS galaxies are again available from the MPA-JHU catalog. To carry out a fair comparison with our high-$z$ blue star-forming galaxies, we restricted to the 287 blue xCOLD~GASS galaxies, with NUV$-$r$<4$; { the stellar mass distribution of the 287 galaxies is shown in Figure~\ref{fig:stellar-mass}[A]}. 252 of these galaxies have CO(1--0) detections, while 35 galaxies have upper limits on the CO(1--0) line luminosity \citep{Saintonge17}.
  The xCOLD~GASS catalogue provides the molecular gas mass of the galaxies, including the mass contribution from helium. We computed the average molecular gas mass of these 287 galaxies, using weights in the average such that the stellar-mass distribution of the xCOLD~GASS galaxies is identical to that in Fig.~\ref{fig:stellar-mass}[C], i.e. identical to that of the GMRT-CAT$z1$ galaxies at $z=1.25-1.45$. For the 35~galaxies with CO(1--0) non-detections, we assume that the molecular gas mass is equal to the $3\sigma$ upper limit on $\Mmol$, provided by the xCOLD~GASS survey. The average molecular gas mass thus obtained is listed in Table~\ref{tab:mass}; {the error on this quantity was obtained from bootstrap resampling with replacement, accounting for the stellar-mass-based weight of each galaxy. }

\section{Results and Discussion}
\label{sec:discuss}

\begin{figure*}
    \centering
    \includegraphics[width=\linewidth]{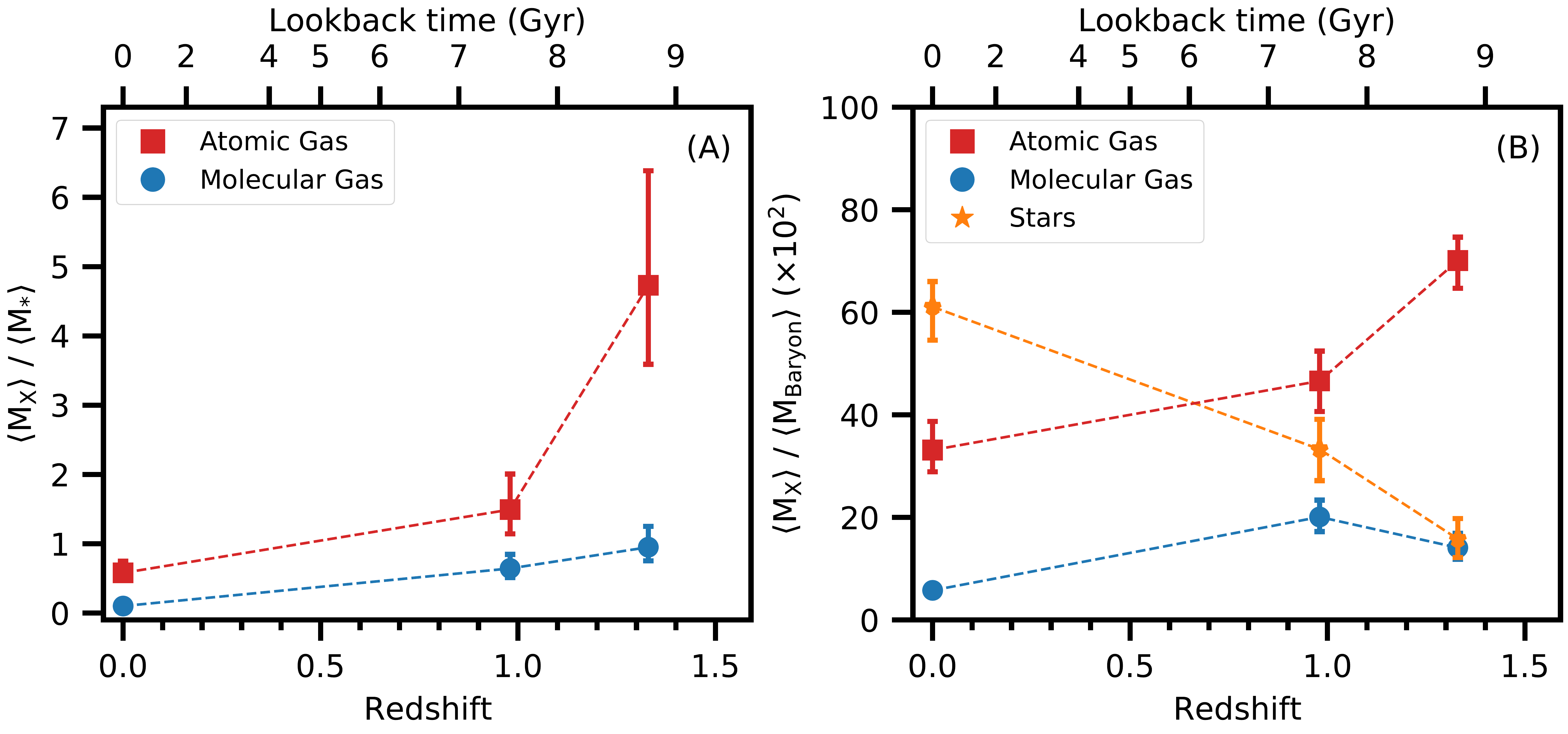}
    \caption{The redshift evolution of (A)~the ratios of the average atomic gas mass and the average molecular gas mass to the average stellar mass, and (B)~the ratios of the average atomic gas mass, the average molecular gas mass, and the average stellar mass to the average baryonic mass. In both panels, all plotted values are for stellar-mass matched samples of galaxies with $\langle\Ms\rangle \approx10^{10}~\Msun$ at $z\approx0$, $z\approx1.0$ and $z\approx1.3$.  The error bars along the y-axis show the $1\sigma$ uncertainties in the estimates. See text for discussion.}
    \label{fig:baryonfraction}
\end{figure*}

\begin{figure*}
    \centering
    \includegraphics[width=0.65\linewidth]{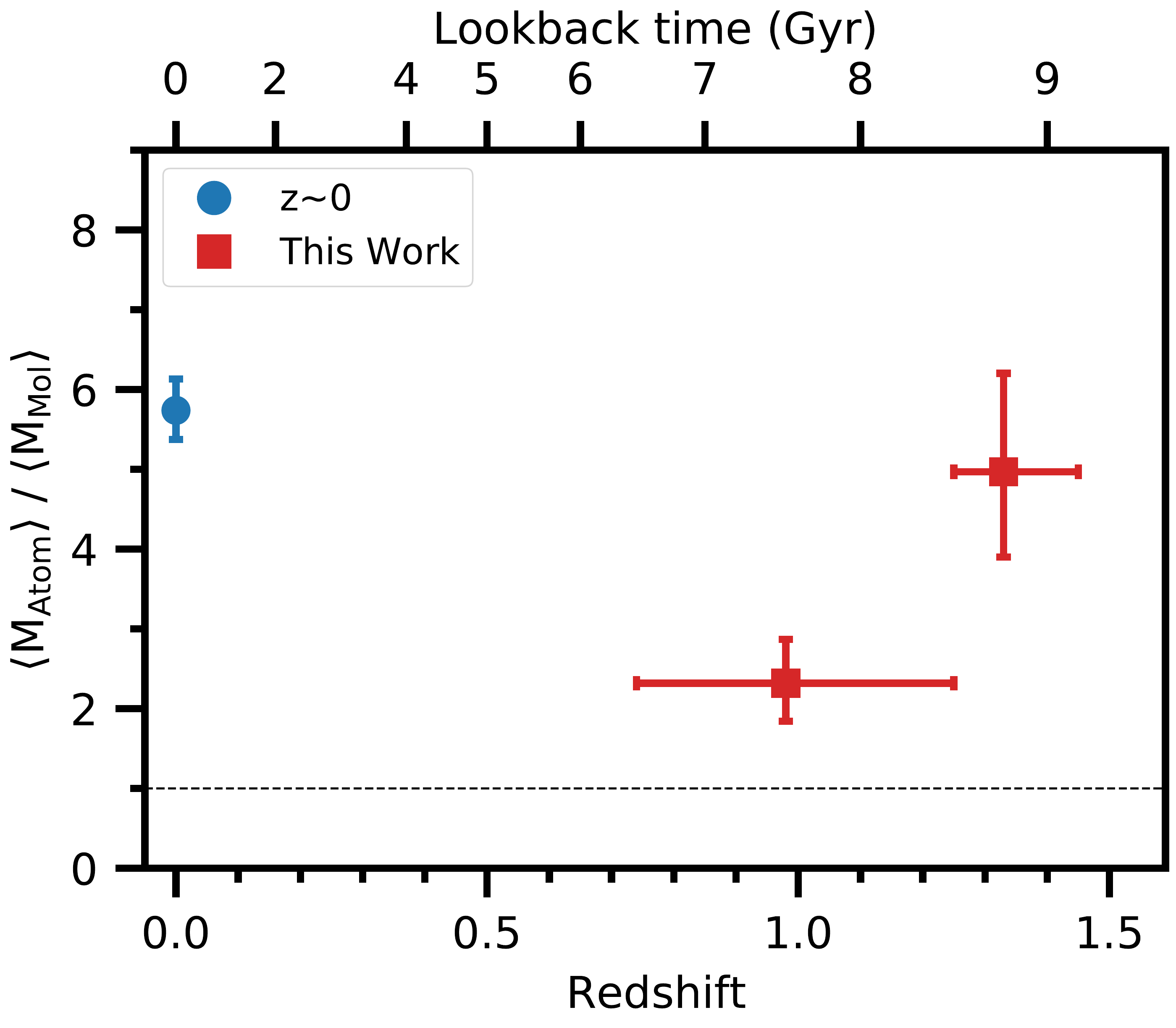}
    \caption{The ratio of the average atomic gas mass to the average molecular gas mass in stellar-mass matched samples of star-forming galaxies at $z\approx0$, $z\approx1.0$, and $z\approx1.3$. The blue circle shows the estimated value of $\langle \Mat \rangle/\langle \Mmol \rangle$ in star-forming galaxies at $z\approx0$ \citep{Saintonge17,Catinella18}. The red squares show estimates of $\langle\Mat\rangle/\langle\Mmol\rangle$ in star-forming galaxies at two redshift intervals $z=0.74-1.25$ and $z=1.25-1.45$. The error bars along the y-axis show the $68.3\%$ confidence intervals on $\langle\Mat\rangle/\langle\Mmol\rangle$. The dashed line indicates $\langle\Mat\rangle/\langle\Mmol\rangle=1$. It is clear that the average atomic gas mass in star-forming galaxies is higher than the average molecular gas mass over the past 9~Gyr.  }
    \label{fig:HIH2}
\end{figure*}

Table~\ref{tab:mass} lists our measurements of the average atomic gas mass, and estimates of the average molecular gas mass, of the GMRT-CAT$z1$ galaxies at $z\approx1.0$ and $z\approx1.3$, along with measurements of the average atomic gas mass and average molecular gas mass of reference samples of blue star-forming galaxies at $z\approx0$. The weights used in the averages ensure that the $z\approx 0$, $z \approx 1$, and $z\approx 1.3$ galaxy samples all have identical stellar-mass distributions, with an average stellar mass of $\langle \Ms \rangle = 10.3 \times 10^{9} \ \Msun$. We note that the listed errors on the $\langle \Mmol \rangle$ values do not include uncertainties in the assumptions (e.g., the value of $\alpha_{\rm CO}$). The uncertainty on $\langle \Ms \rangle$ is assumed to be 0.1~dex, based on comparisons between stellar-mass estimates for the same galaxies using different methods and assumption \citep[e.g.][]{Stefanon17}. The last row of the table combines the estimates of the average atomic gas mass, average molecular gas mass, and the average stellar mass to infer the average total baryonic mass of star-forming galaxies at $z\approx0$, $z\approx1.0$, and $z\approx1.3$. The average baryonic mass,  $\langle\MBary\rangle$, for each redshift interval was estimated using $\langle\MBary\rangle=\langle\Ms\rangle+\langle\Mat\rangle+\langle\Mmol\rangle$. 

Table~\ref{tab:ratios} lists the ratios of the average atomic gas, molecular gas, and stellar masses relative to the average stellar mass and the average baryonic mass, as well as the ratio of the average atomic gas mass to the average molecular mass. We estimated the errors on each ratio via Monte Carlo simulations in which we obtained a large number of realisations of the ratio by drawing pairs of values for the average masses in the numerator and the denominator from Gaussian distributions of the two quantities, with the same means and standard derivations as the estimates of the average masses.

It is clear from Table~\ref{tab:ratios} that the baryonic composition of star-forming galaxies shows dramatic evolution over the last $\approx 9$~Gyr, from $z \approx 1.3$ to $z = 0$. 
Figure~\ref{fig:baryonfraction}(A) plots the ratios of the average atomic gas and molecular gas masses (red squares and blue circles, respectively) to the average stellar mass versus redshift.  The figure shows that that both $\langle\Mmol\rangle/\langle\Ms\rangle$ and $\langle\Mat\rangle/\langle\Ms\rangle$ decrease by roughly an order of magnitude from $z \approx 1.3$ to $z\approx0$. However, the nature of the decline is very different in the atomic and the molecular components. The ratio  $\langle\Mat\rangle/\langle\Ms\rangle$ drops steeply, by a factor of $\approx3.3$, in the $\approx1$~Gyr period between $z\approx1.3$ and $z\approx1.0$, and then falls gradually, by a factor of $\approx2.6$, over the $\approx7$~Gyr between $z\approx1.0$ and $z\approx0.0$. Conversely, the ratio $\langle\Mmol\rangle/\langle\Ms\rangle$ falls by a factor of only $\approx1.5$ between $z\approx1.3$ and $z\approx1.0$, but then drops by a factor of $\approx6.3$ between $z\approx1$ and $z\approx0$. The rapid decline in the atomic gas mass of galaxies between $z\approx1.3$ and $z\approx1.0$, towards the end of the epoch of galaxy assembly, indicates insufficient accretion of gas from the CGM \citepalias{Chowdhury22a}; this is the likely cause for the decline in the star-formation activity of the Universe at $z\lesssim1$. Further, it is clear from Fig.~\ref{fig:baryonfraction}(A) that the atomic gas mass is significantly higher than the stellar mass, by a factor of $\approx 4.4$, at $z \approx 1.3$, during the epoch of peak star-formation activity, $z \approx 1-3$, in the universe \citep{Madau14}.

Figure~\ref{fig:baryonfraction}(B) plots the redshift evolution of the stellar, atomic gas, and molecular gas fractions of the baryonic mass. The three main baryonic components of galaxies show very different behaviours. In the local Universe, it is clear that stars dominate the baryonic content of star-forming galaxies with $\langle\Ms\rangle\approx10^{10}~\Msun$, constituting $\approx60\%$ of the baryonic mass. However, the fraction of baryons in stars decreases with increasing redshift: stars make up only $16\%$ of the baryonic mass in such galaxies at $z\approx1.3$. Conversely, Figure~\ref{fig:baryonfraction}(B) shows that the contribution of both atomic gas and molecular gas to the total baryonic mass of star-forming galaxies with $\langle\Ms\rangle\approx10^{10}~\Msun$ is significantly higher at $z \gtrsim 1$ than in the local universe. The contribution of atomic gas to the baryonic mass increases from $\approx33\%$ at $z\approx0$ to $\approx47\%$ at $z\approx1$, and then to $\approx70\%$ at $z\approx1.3$. For the molecular component, we find that $\langle\Mmol\rangle/\langle\MBary\rangle$ increases from $\approx6\%$ at $z\approx0$ to $\approx20\%$ at $z\approx1.0$, and then flattens, with $\langle\Mmol\rangle/\langle\MBary\rangle \approx 14\%$  at $z\approx1.3$. Overall, Figure~\ref{fig:baryonfraction}(B) shows that the neutral-gas fraction of the baryonic mass of star-forming galaxies with $\langle\Ms\rangle\approx10^{10}~\Msun$ at $z\gtrsim1$ is significantly higher than at $z\approx0$. Neutral gas makes up $\approx 84$\% of the baryonic mass of star-forming galaxies at $z \approx 1.3$, with atomic gas constituting $\approx70\%$ of the baryonic mass.

Finally, the values of the ratio of the average atomic gas mass to the average molecular gas mass in star-forming galaxies with $\langle\Ms\rangle\approx10^{10}~\Msun$ at $z\approx0$, $z\approx1.0$ and $z\approx1.3$ are listed in Table~\ref{tab:mass} and plotted against redshift in Figure~\ref{fig:HIH2}. We find that $\langle \Mat \rangle/\langle \Mmol \rangle$ decreases from $5.74^{+0.39}_{-0.37}$ in the local Universe to $2.32^{+0.54}_{-0.47}$ at $z\approx1$. Interestingly, however, we find that the ratio shows evidence for an increase at higher redshifts, $z > 1$, with  $\langle\Mat\rangle/\langle\Mmol\rangle=5.0^{+1.2}_{-1.0}$  for galaxies with $\langle\Ms\rangle\approx10^{10} \ \Msun$ at $z\approx1.3$. Atomic gas thus clearly dominates the cold-gas content of star-forming galaxies at $z\approx1.3$.

A possible source of error in our $\langle \Mmol \rangle$ estimates for the CAT$z$1 galaxies lies in the extrapolation of Equation~\ref{eqn:tacconi20} to galaxies with stellar masses $\approx 10^9 \ \Msun$ at $z \gtrsim 0.7$, a regime that is not tightly constrained by current data \citep{Tacconi20}. 
However, {the average $\langle \Mmol \rangle$ value is dominated by galaxies with $\Ms>10^{10}~\Msun$}. Thus, even if we assume that the molecular gas masses of all galaxies with $\Ms < 10^{10}~\Msun$ are systematically higher by a factor of $\approx 5$ than the values obtained from Equation~\ref{eqn:tacconi20}, this would only increase $\langle \Mmol \rangle$ at $z \approx 1.3$ by a factor of $\approx 2$, yielding $\langle \Mat \rangle/\langle \Mmol \rangle \approx 2.5$  and $\langle \Mat \rangle/\langle \MBary \rangle \approx 61\% $. Another possible source of error in the $\langle \Mmol\rangle$ estimates stems from the uncertainties in the assumptions (e.g. the value of $\alpha_\textrm{CO}$, the dust-to-gas ratio, etc) made when originally determining the molecular gas masses that were used to obtain  Equation~\ref{eqn:tacconi20}; such uncertainties are expected to be $\approx \pm0.25$~dex \citep{Tacconi20}. However, even assuming that the ``true'' molecular gas masses are all 0.25~dex higher than those inferred from Equation~\ref{eqn:tacconi20}, we find that our estimate of $\langle \Mat \rangle/\langle \Mmol \rangle$ at $z \approx 1.3$ would decrease to $2.8$, and of $\langle \Mat \rangle/\langle \Mmol \rangle$ to $\approx 63\%$. Thus, our conclusion that atomic gas dominates the cold-gas content of star-forming galaxies at $z \approx 1.3$ appears to be robust against even relatively large uncertainties in the average molecular gas mass of high-$z$ galaxies.

Figure~\ref{fig:HIH2} shows that atomic gas is the dominant component of the cold ISM of galaxies at both $z\approx0$ and $z\gtrsim1$. Spatially-resolved \hii\ and CO studies in nearby galaxies find that the \htwo\ arises in the inner, star-forming, regions of galaxies, while the \hi\ is much more extended, extending to radii of tens of kpc \citep[e.g.][]{Leroy08}. Indeed, the molecular gas mass dominates the ISM in the central regions of spiral galaxies at $z\approx0$, with the transition from an \hi-dominated ISM to an \htwo-dominated ISM occurring at approximately half the optical radius, at a characteristic gas surface density of $\approx 14~\Msun \textrm{pc}^{-2}$ \citep{Leroy08}. For high-$z$ galaxies, CO emission in star-forming galaxies has been found to have a half-light radius of $\lesssim 10$~kpc, similar to the size of the star-forming regions \citep[e.g.][]{Tacconi13,Bolatto15}. Conversely, we find that the average \hii\ emission from star-forming galaxies at $z\approx1$ is resolved for spatial resolutions  $< 90$~kpc \citepalias{Chowdhury22b}. {It thus appears that the \htwo\ in high-$z$ star-forming galaxies is also restricted to the central high-density regions while the \hi\ extends out to tens of kpc in a significant fraction of such galaxies.}

In this \emph{Letter}, we have shown that the average atomic gas mass of star-forming galaxies with $\langle\Ms\rangle\approx10^{10}~\Msun$ is comparable to the average stellar mass at $z \approx 1$, and is significantly larger than both the average stellar mass and the average molecular gas mass at $z \approx 1.3$. We find that $\approx70\%$ of the baryonic mass of star-forming galaxies  with $\langle\Ms\rangle\approx10^{10}~\Msun$  at $z\approx1.3$ is in atomic gas. Our results thus demonstrate that atomic gas dominates the baryonic content of star-forming galaxies at $z \approx 1.3$, during the epoch of peak star-formation activity in the universe.

	\begin{acknowledgments}
	We thank the staff of the GMRT who have made these observations possible. The GMRT is run by the National Centre for Radio Astrophysics of the Tata Institute of Fundamental Research. We thank an anonymous referee for a very constructive report that helped increase the clarity of the paper. NK acknowledges support from the Department of Science and Technology via a Swarnajayanti Fellowship (DST/SJF/PSA-01/2012-13). AC, NK, $\&$ JNC also acknowledge the Department of Atomic Energy for funding support, under project 12-R\&D-TFR-5.02-0700. 
	\end{acknowledgments}
    \software{ numpy \citep{harris2020array};    matplotlib \citep{Hunter:2007}}
      
    \bibliography{bibliography.bib}

\bibliographystyle{aasjournal}

\end{document}